\documentstyle[aps,prl,twocolumn,psfig]{revtex}
\begin{document}
\preprint{\vbox{}}
\draft
\title{Electroweak Model Independent Tests for SU(3) Symmetry 
\\in Hadronic B Decays}
\author{Xiao-Gang He, Jenq-Yuan Leou and Chung-Yi Wu}
\address{
Physics Department, National Taiwan University, 
Taipei, Taiwan 10764, R.O.C.
}
\date{ July, 2000}
\maketitle
\begin{abstract}
We study effects of new physics beyond the Standard Model on SU(3)
symmetry in charmless hadronic two body B decays. It is found that 
several equalities for some of the decay amplitudes, such as
$A(B_d (B_u) \to$ $ \pi^+\pi^-,$ $\pi^+ K^- (\pi^- \bar K^0))$ $ = 
$ $A(B_s \to $ $K^+ \pi^-, $ $K^- K^+ $ $(K^0 \bar K^0))$,
$A(B_d \to $ $\pi^+\rho^-, $ $\pi^- \rho^+,$ 
$ K^-\rho^+, $ $\pi^+ K^{*-} )$ 
$ = $ $A(B_s \to $ $K^+ \rho^-,$ 
$ \pi^- K^{*+},$ $ K^- K^{*+}, $ $K^+ K^{*-})$,
$A(B_d (B_u) \to $ $\rho^+\rho^-,$ 
$\rho^+ K^{*-}$ $ (\rho^- \bar K^{*0}))$ $ = 
$ $A(B_s \to$ $ K^{*+} \rho^-,$ 
$ K^{*-} K^{*+}$ $ (K^{*0} \bar K^{*0}))$,
predicted by SU(3) symmetry in the SM are not 
affected by new physics. These relations provide important 
electroweak model independent tests for SU(3) 
symmetry in B decays. 

\end{abstract}


\preprint{\vbox{\hbox{}}}


The calculations of hadronic B decays are difficult due to our 
poor understanding of
QCD hadronic physics. Model calculations can account 
for some of the 
measured B decays, but not all of them\cite{1,2}. 
The results from model calculations
are, in any case, far from the desired accuracy needed to test 
the Standard Model (SM) and models beyond. 
To overcome some of the difficulties, 
several groups have proposed to 
use SU(3) symmetry to study charmless hadronic two body 
B decays and have obtained some interesting 
results\cite{3,4,5,6,7}.
At present SU(3) symmetry is not well tested, although 
it has been shown
to be consistent with data from $B_u \to D^0 \pi^- (K^-)$ 
decays\cite{5}. In order to have
a good understanding of the underlying theory of B 
decays using information extracted from
SU(3) symmetry considerations, it is important to 
know to what precision the SU(3) 
symmetry is valid. To 
achieve this, one
should not only to obtain testable relations predicted by 
SU(3) symmetry in the SM, 
but also to make sure that the relations are not modified 
by new physics beyond the
SM for electroweak interactions. 
In this paper we study effects of new physics beyond the 
SM on SU(3) symmetry predictions
in some of the charmless hadronic two body B decays into 
SU(3) octet 
pseudoscalars $P$ and/or vectors $V$. 
We find that, indeed, it is possible to test 
SU(3) symmetry in a way which is independent of models 
for electroweak interactions. 

We find some equalities for hadronic B decays using
SU(3) symmetry when small
annihilation contributions are neglected. 
These are the following relations for $B \to PP$ modes

\begin{eqnarray}
&&A(B_d (B_u) \to \pi^+\pi^-,\pi^+ K^- (\pi^- \bar K^0))
\nonumber\\
&& = 
A(B_s \to K^+ \pi^-,K^- K^+ (K^0 \bar K^0)),
\label{pp}
\end{eqnarray}
and the following relations for $B\to PV$ and $B\to VV$ modes

\begin{eqnarray}
&&A(B_d \to \pi^+\rho^-, \pi^- \rho^+, K^-\rho^+, 
\pi^+ K^{*-} )\nonumber\\
&& = 
A(B_s \to K^+ \rho^-, \pi^- K^{*+}, K^- K^{*+}, K^+ K^{*-}), 
\nonumber\\
&&A(B_d (B_u) \to \rho^+\rho^-,\rho^+ K^{*-} 
(\rho^- \bar K^{*0}))\nonumber\\
&& = 
A(B_s \to K^{*+} \rho^-, K^{*-} K^{*+} (K^{*0} \bar K^{*0})).
\label{pv}
\end{eqnarray}
Further we find that these equalities are not 
affected by new physics 
beyond the SM.
We note that relations in eqs. (\ref{pp}) and (\ref{pv}) 
always involve charmless hadronic 
$B_s$ decays which have
not been measured. 
However these decay modes have relatively 
large branching ratios ($10^{-5}\sim 10^{-6}$) 
from model calculations and are expected to be 
measured at CDF, D0, HERA-B, LHC-B 
and BTeV. When all 
related modes are measured SU(3)
symmetry will be tested in an electroweak model independent 
way. In the following we 
provide more details.

In the SM the quark level effective Hamiltonian for 
charmless hadronic B decays, including QCD
corrections, can be written as

\begin{eqnarray}
H^q_{eff} &=& {G_F \over \sqrt{2}} [
V_{ub}V_{uq}^*(c_1 O_1 + c_2 O_2)\nonumber\\
&-&\sum_{3}^{12}(c_i^{uc}
+V_{tb}V_{tq}^* c_i^{tc}) O_i^q ],\nonumber
\end{eqnarray}
where $i$ is summed over from 3 to 12, 
$c_{1,2}$ and $c_i^{jk} = c_i^j - c_i^k$ are the 
Wilson coefficients which have 
been evaluated in various renormalization schemes\cite{8}. 
Here $j,\;k$
indicate the internal quarks for loop induced operators. 
$q$ can be $d$ or $s$ 
depending on whether the decays are $\Delta S =0$ or 
$\Delta S = 1$. The specific
form of the operators can be found in Ref.\cite{8}. Since 
we will only need to know their SU(3) structures, we  
will suppress Lorentz structure and treat
the Fierz transformed and un-transformed forms as the 
same in our later discussions.
 
The operators $O_{3,4,5,6,11}$ all have a simple SU(3)
structure and transform as
$\bar 3$. The complication comes from the fact that 
$O_{1,2,7,8,9,10,12}$ are not a single SU(3) 
irreducible representation. 
They
contain two $\bar 3$'s, one 6 and one $\overline{15}$. 
Although $O_{7,8,9,10}$ are electroweak type with small 
Wilson coefficients, they
are important for B decays and must be kept\cite{9}.
For illustration we give the detailed decomposition of 
$O_2=\bar u b \bar q u$ for $q=d$ and have

\begin{eqnarray}
\bar u b \bar d u 
&=& {1\over 8} \left \{ 3[\bar u b\bar d u + \bar d b 
\bar d d + \bar s b \bar d s]_{\bar 3}
-[\bar d b(\bar u u+\bar d d+\bar s s)]_{\bar 3'}\right .
\nonumber\\
&+& 2[\bar u b \bar d u - \bar d b \bar u u + \bar d b 
\bar s s - \bar s b \bar d s]_6\nonumber\\
&+&\left . [3\bar u b \bar d u +3 \bar d b \bar u u - 2 
\bar d b \bar d d - 
\bar d b \bar s s -\bar s b \bar d s]_{\overline{15}}
\right \}\nonumber\\
&=& {1\over 8} [3 H(\bar 3) - H(\bar 3') + 2H(6) + 
H(\overline{15})],
\end{eqnarray}
where $H(i)$ are matrices in SU(3) flavor space. 
With the identification of
$u=1$, $d=2$ and $s=3$, the non-zero entries of 
$H(i)$ are given by\cite{5}

\begin{eqnarray}
&&H(\bar 3^{(\prime)}) ^2 =1,\;
H(6)^{12}_1 = H(6)^{23}_3 =1,\; 
\nonumber\\
&&
H(\overline{15})^{12}_1 =3,\; 
H(\overline{15})^{22}_2 = -2,\; 
H(\overline{15})^{32}_3 = -1.
\end{eqnarray}
$H(6)$ and $H(\overline{15})$ are
 anti-symmetric and symmetric in exchaning the upper two
indices, respectively.

For $q=s$ case, we have\cite{5}

\begin{eqnarray}
&&H(\bar 3^{(\prime)}) ^2 =1,\;
H(6)^{13}_1 = H(6)^{32}_2 = 1,\;\nonumber\\
&&H(\overline{15})^{13}_1 =3,\;H(\overline{15})^{33}_3 = -2,\;
H(\overline{15})^{32}_2 = -1.
\end{eqnarray}
The $H(i)$ matrices for other operators 
can be obtained in a similar way. 

At the hadronic level, the decay amplitudes can be written as

\begin{eqnarray}
A(B\to i j) 
= V_{ub}V_{uq}^* T^{SM} 
+ V_{tb}V_{tq}^*P^{SM}.
\end{eqnarray}
Here $T^{SM}$ and $P^{SM}$ both have $\bar 3$, $6$ and 
$\overline{15}$ components through 
operators associated with
$c_{1,2}$, $c_i^{uc}$ and $c_i^{tc}$. Since we only 
concern the SU(3) structure, the
detailed coefficients are not important, that is, 
difference operators having the same SU(3)
irreducible representations can be combined together 
and denoted by certain SU(3) invariant
amplitudes. The amplitudes $T^{SM}$ can be written in 
the following form for 
$B\to PP$ decays\cite{5},

\begin{eqnarray}
&&T^{SM} = A^T_{\bar 3} B_i H(\bar 3)^i (M_l^k M_k^l) 
+ C^T_{\bar 3}B_i M^i_k M^k_j H(\bar 3)^j
\nonumber\\
&&+A^T_{6} B_i H(6)^{ij}_k (M_j^l M_l^k) + 
C^T_{6}B_i M^i_jH(6)^{jk}_l M^l_k\nonumber\\
&&+A^T_{\overline{15}} B_i H(\overline{15})^{ij}_k (M_j^l M_l^k) 
+ C^T_{\overline{15}}
B_i M^i_jH(\overline{15})^{jk}_l M^l_k,
\label{amplitude}
\end{eqnarray}
where $M^i_j$ is the pseudoscalar octet $P$. 
$B_i$ is the B-meson SU(3) triplet $(B_u,\;B_d,\;B_s)$. 
In the case for $B\to PP$ decays $A_6$ and $C_6$ always 
appear together 
in the form $C_6 - A_6$\cite{5}. We will eliminate $A_6$ in
the expressions.
$P^{SM}$ can be obtained similarly.

The amplitudes $A_i$ correspond to annihilation 
contributions which
can be seen from eq. (\ref{amplitude}) where $B_i$ is contracted 
with one of the indices
in $H$ matrices. These contributions are much smaller than the 
amplitudes $C_i$ from model
calculations\cite{2,5}. The smallness of these annihilation amplitudes 
can be tested by measuring
the branching ratios for  
$B_d \to K^+ K^-$, $B_s \to \pi^+\pi^-, \pi^0\pi^0$ 
which are proportional
to $A_{\bar 3} + A_{\overline{15}}$\cite{4,5}.
We will work with the assumption that annihilation contributions 
are small and can be neglected. 
Future
experiments will decide if this assumption is valid\cite{4,5,6}. 
Expanding eq. (\ref{amplitude})
one obtains the decay amplitudes in 
terms of $A_i$ and $C_i$. The relevant decay amplitudes are given by

\begin{eqnarray}
&&T^{SM}_{B_d \to \pi^+\pi^-, B_s \to K^- K^+}= 2A^T_{\bar 3} 
+ A^T_{\overline{15}} + 
C^T_{\bar 3} + C^T_6 + 3 C_{\overline{15}}^T,\nonumber\\
&&T^{SM}_{B_s \to K^+ \pi^-, B_d \to \pi^+ K^-}= 
- A^T_{\overline{15}} + C^T_{\bar 3} + C^T_6 + 3 C_{\overline{15}}^T
;\nonumber\\
&&T^{SM}_{B_u \to \pi^- \bar K^0}
 = 3 A^T_{\overline{15}} 
+ C^T_{\bar 3} 
- C^T_6 - C^T_{\overline{15}},\nonumber\\
&&T^{SM}_{B_s \to K^0 \bar K^0}
 =2 A^T_{\bar 3}-3 A^T_{\overline{15}} 
+ C^T_{\bar 3} 
- C^T_6 - C^T_{\overline{15}}.
\label{ppp}
\end{eqnarray}
In naive quark diagram analysis, when annihilation contributions are 
neglected, $B_u(B_s) \to \pi^- \bar K^0 (K^- \bar K^0)$ do not have
contributions from $O_{1,2}$\cite{3,4}. 
This is not true when going beyond
naive quark diagram analysis and need to be tested\cite{5}. Our 
results, however, do not rely on whether they are zero or not.
 
For $B\to VV$, the decay amplitudes can be obtained from $B\to PP$ 
by a simple replacement of the
corresponding final states. 
The decay amplitudes for
$B\to P V$ are more complicated because the fact that there are two terms,
except for $A_3$, for 
each of the terms in eq. (\ref{amplitude}). For example,
the term corresponding to $C^T_{\bar 3} B_i M^i_k M^k_jH(\bar 3)^j$ 
becomes,
$C^V_{\bar 3} B_iV^i_k M^k_j H(\bar 3)^j$ and $C^M_{\bar 3} 
B_i M^i_k V^k_j H(\bar 3)^j$. Although there are relations between 
$A_6^{V,M}$ and $C_6^{V,M}$, they do not always appear 
together as $C_6 -A_6$ like for
$B\to PP$. We will need to keep all of them.
The details for the whole amplitudes can be found in Ref.\cite{6}.
The smallness of annihilation contributions
for $B\to VV$ and $B\to PV$ can, again, be tested
by measuring some pure annihilation decays such as
$B_d \to K^{*+} K^{*-}, K^+ K^{*-}, K^- K^{*+}$, $B_s \to \rho^+ \rho^-, 
\rho^0 \rho^0, \pi^+ \rho^-, \pi^-\rho^+, \pi^0 \rho^0$.

From eq. (\ref{ppp}) and Tables in Ref.\cite{6}, we find that when 
annihilation contributions are
neglected (setting all $A_i$ to zero), 
the equalities in eqs. (\ref{pp}) and (\ref{pv}) 
hold in the SM. 
These relations can be used to test whether SU(3) is a good 
symmetry for
B decays in the SM. 

The relations in eqs. (\ref{pp}) and (\ref{pv}) 
hold in SU(3) symmetry limit. 
There are SU(3) breaking effects. These
include differences in phase space for $B_{u,d}$ and $B_s$ decays, and 
also in the decay amplitudes. 
It is not possible to reliably calculate the breaking 
effects in the decay amplitudes
at present. 
Factorization approximation gives, for example, 
$A(B_d (B_u) \to $ $\pi^+\pi^-,$ $ \pi^+ K^- $ $(\pi^- \bar K^0))$ $
=$ 
$(F_0^{B_d \pi}/ F_0^{B_s K})$ $
A(B_s \to$ $ K^+ \pi^-,$ $ K^- K^+$ $ (K^0 \bar K^0))$,
where $F^{B_i j}_0$ are transition form factors. 
Model calculations indicate that the ratio $F^{B_{u,d} 
\pi}_0/F^{B_s K}_0$ is
close to one. Deviation from one for this ratio 
would be an indication of SU(3) 
breaking in B decays. 
%
Estimates of SU(3) breaking effects for other 
decays can be obtained in a similar way. As already mentioned that
accurate theoretical calculations are very difficult to carry out, 
we, therefore,
will not attempt to obtain precise theoretical predictions 
of SU(3) symmetry 
breaking effects, but 
to study if the relations in eqs. (\ref{pp}) and (\ref{pv}) 
are modified by new physics
beyond the SM and further to study if SU(3) symmetry and its breaking 
can be determined in 
an electroweak model independent way.

In the presence of new physics beyond the SM, there are new operators 
in addition to the ones
already present in the SM. As long as SU(3) structure is concerned, 
there are two types
of new operators which can appear at the four quark level for 
charmless hadronic $\Delta S = 0$ and  $\Delta S = 1$ B decays. 
These operators are

\begin{eqnarray}
O_{\bar q \bar d d} = \bar q b \bar d d,\;\;\;\;O_{\bar q \bar s s} 
= \bar q b \bar s s.
\end{eqnarray}
These operators can naturally appear in extensions of the SM. 
For example the term $U_RD_R D_R$ in 
R-Parity violating supersymmetric models\cite{10} can induce 
$ O_{\bar q \bar d d, \bar q \bar s s}$ by 
exchanging s-up-quarks with sizeable contributions to B decays. 
In the same model exchanging
s-down-quarks can also generate $\bar u b \bar q u$ type of 
operators which has the same SU(3)
structure as the operators $O_{1,2}$ in the SM, but with difference 
Lorentz structure.

One may wonder if considering operators just up to dimension six 
are sufficient.
Higher order operators may have more complicated
structure due to higher SU(3) irreducible representations. 
However, because exchange of gluons, soft or hard, will 
not change the SU(3) structure, the 
contributions of higher order operators with different SU(3) 
structures will 
have additional suppression
factors from loop integrals or additional propagators of 
electroweak types in a model and can be 
neglected. Thus considering operators up to dimension six 
is sufficient for our purpose.

The operators $O_{\bar d\bar d d}$ and $O_{\bar s \bar s s}$ 
contain only $\bar 3^{(\prime)}$ and
$\overline{15}$. The non-zero entries of $H(\bar 3^{(\prime)})$ 
are the same as the corresponding ones in the SM with an 
appropriate normalizations. The non-zero entries of 
$H(\overline{15})$ are given by

\begin{eqnarray}
&&H^{\bar d\bar d d}(\overline{15})^{12}_1
=H^{\bar d\bar d d}(\overline{15})^{23}_3=-2,\;
H^{\bar d\bar dd}(\overline{15})^{22}_2 = 4,\;\; 
\nonumber\\
&&H^{\bar s\bar ss}(\overline{15})^{13}_1
=H^{\bar s\bar ss}(\overline{15})^{23}_2=-2,\;
H^{\bar s\bar ss}(\overline{15})^{33}_3 = 4,
\end{eqnarray}

The operators $O_{\bar s \bar dd}$ and $O_{\bar d \bar s s}$ 
contain $\bar 3^{(\prime)}$, $6$ and
$\overline{15}$. The non-zero entries of $H(\bar 3^{(\prime)})$ 
can again be normalized to 
be the same as the corresponding
ones in the SM. The non-zero entries of the $6$ and 
$\overline{15}$ are

\begin{eqnarray}
&&H^{\bar d \bar s s}(6)^{23}_3 = H^{\bar d \bar s s}(6)^{12}_1 
= 1,\nonumber\\
&&H^{\bar d \bar s s}(\overline{15})^{23}_3 =  3,\;
H^{\bar d \bar s s}(\overline{15})^{22}_2 = -2,\; 
H^{\bar d \bar s s}(\overline{15})^{12}_1 = -1,\nonumber\\
&&H^{\bar s \bar dd}(6)^{13}_1 = H^{\bar s \bar dd}(6)^{32}_2 = 
 1,\nonumber\\
&&H^{\bar s \bar dd}(\overline{15})^{23}_2 =3,\;
H^{\bar s \bar dd}(\overline{15})^{33}_3 = -2,\;
H^{\bar s \bar dd}(\overline{15})^{13}_1 = -1.
\end{eqnarray}

With these new operators the charmless hadronic 
B decay amplitudes will be modified.
Normalizing to the SM amplitudes, we can write 
the total amplitudes as

\begin{eqnarray}
A(B\to i j)&=& V_{ub}V_{uq}^* T^{SM} 
+ V_{tb}V_{tq}^*P^{SM}
\nonumber\\
&+&a^{\bar q \bar u u} P^{\bar q \bar u u}
+a^{\bar q \bar dd} P^{\bar q \bar dd}
+a^{\bar q \bar ss} P^{\bar q \bar ss},
\end{eqnarray}
where $a^i$ indicate the coefficients due to new 
physics beyond the SM, and
$P^{\bar q \bar ll} = <ij|O_{\bar q \bar l l}|B>$. 
Here we have also included the contributions
from operators of the form $O_{\bar q u u}$ which 
have the same SU(3) structure as
$O_{1,2}$ in the SM, but are due to new physics 
and also may have difference Lorentz structures.

Following the same procedure as for the SM discussed 
before, one can obtain the decay amplitudes
in terms of the SU(3) invariant amplitudes. 
The decay amplitudes due to the new operators for 
the relevant $B\to PP$ modes are given as below.

For $\Delta S = 0$ decay modes, we have

\begin{eqnarray}
&&P^{\bar d \bar dd}_{B_d \to \pi^+\pi^-} 
=  2A_{\bar 3}^{\bar d \bar d d} 
+ 2 A_{\overline{15}}^{\bar d \bar dd}
+C_{\bar 3}^{\bar d \bar d d} 
- 2C_{\overline{15}}^{\bar d \bar dd},\nonumber\\
&&P^{\bar d \bar dd}_{B_s \to  K^+\pi^-} =  
-2 A_{\overline{15}}^{\bar d \bar dd}
+C_{\bar 3}^{\bar d\bar dd}
- 2C_{\overline{15}}^{\bar d \bar dd};\nonumber\\
&&P^{\bar d \bar s s}_{B_d \to \pi^+\pi^-} 
= 2A_{\bar 3}^{\bar d \bar  s s} 
- 3 A_{\overline{15}}^{\bar d \bar s s}
+C_{\bar 3}^{\bar d \bar ss}
+ C_6^{\bar d \bar s s}
- C_{\overline{15}}^{\bar d \bar s s},\nonumber\\
&&P^{\bar d \bar s s}_{B_s \to  K^+\pi^-} 
=  3 A_{\overline{15}}^{\bar d \bar s s}
+C_{\bar 3}^{\bar d \bar ss}
+ C_6^{\bar d \bar s s} 
- C_{\overline{15}}^{\bar d \bar s s}.
\end{eqnarray}

For the two pairs of $\Delta S = 1$ decay modes, 
we have

\begin{eqnarray}
&&P^{\bar s \bar ss}_{B_d \to \pi^+ K^-} 
=  
- 2 A_{\overline{15}}^{\bar s \bar ss}
+C_{\bar 3}^{\bar s \bar s s}
- 2 C_{\overline{15}}^{\bar s \bar ss},\nonumber\\
&&P^{\bar s \bar ss}_{B_s \to  K^+ K^-} = 
2A_{\bar 3}^{\bar s \bar s s} +
2 A_{\overline{15}}^{\bar s \bar ss}
+C_{\bar 3}^{\bar s \bar s s}
- 2 C_{\overline{15}}^{\bar s \bar ss};\nonumber\\
&&P^{\bar s \bar dd}_{B_d \to \pi^+ K^-} 
=   3 A_{\overline{15}}^{\bar s \bar dd}
+C_{\bar 3}^{\bar s \bar d d}
+ C_6^{\bar s \bar dd}
- C_{\overline{15}}^{\bar s \bar dd},\nonumber\\
&&P^{\bar s \bar dd}_{B_s \to  K^+ K^-} 
= 2A_{\bar 3}^{\bar s \bar d d}
- 3 A_{\overline{15}}^{\bar s \bar dd}
+C_{\bar 3}^{\bar s \bar d d}
+ C_6^{\bar s \bar dd} 
- C_{\overline{15}}^{\bar s \bar dd};
\end{eqnarray}
and

\begin{eqnarray}
&&P^{\bar s \bar ss}_{B_u \to \pi^- \bar K^0} 
= - 2A_{\overline{15}}^{\bar s \bar ss}
+C_{\bar 3}^{\bar s \bar s s}
-2 C_{\overline{15}}^{\bar s \bar ss},\nonumber\\
&&P^{\bar s \bar ss}_{B_s \to  K^0 \bar K^0} =
2A_{\bar 3}^{\bar s \bar s s}  
+2 A_{\overline{15}}^{\bar s \bar ss}
+C_{\bar 3}^{\bar s\bar ss}
-2 C_{\overline{15}}^{\bar s \bar ss};\nonumber\\
&&P^{\bar s \bar dd}_{B_u \to \pi^- \bar K^0} 
=  - A_{\overline{15}}^{\bar s \bar dd}
+C_{\bar 3}^{\bar s \bar d d}
- C_6^{\bar s \bar dd}
+3 C_{\overline{15}}^{\bar s \bar dd},\nonumber\\
&&P^{\bar s \bar dd}_{B_s \to  K^0 \bar K^0} =
2A_{\bar 3}^{\bar s \bar d d}
+ A_{\overline{15}}^{\bar s \bar dd}
+C_{\bar 3}^{\bar s \bar d d}
- C_6^{\bar s \bar dd} 
+3 C_{\overline{15}}^{\bar s \bar dd}.
\end{eqnarray}
From the above expressions for 
$P^{\bar q \bar dd, \bar q \bar s s}$, 
we clearly see that when annihilation
contributions are neglected the equalities in eq. (\ref{pp}) hold.
In fact the new operators have zero contributions from $C_i$
to the above amplitudes, except 
$B_{u}(B_s) \to \pi^- \bar K^0 (K^0 \bar K^0)$, 
in the naive quark diagram analysis. Re-scattering effects
may generate non-zero contributions. Our results are, however,
independent from whether these contributions are large. This
also applies to related $B\to VV, PV$ modes.
  
The decay amplitudes for $B\to VV$ can be obtained, again, by a 
simple replacement of the corresponding final states.
The amplitudes for $B\to PV$ are given in Table 1. 
We can see that the SU(3) predictions of the relations in
eq. (\ref{pv}) are not modified if annihilation amplitudes 
are neglected.
We conclude that the relations in eqs. (\ref{pp}) and (\ref{pv}) 
are independent of new physics beyond the SM when small annihilation
contributions are neglected. 
Measurements of these relations provide true tests of SU(3) 
symmetry in charmless
hadronic two body B decays.
There are also other interesting relations, 
with some of them depending on
electroweak model which can be tested experimentally. 
A more detailed study of related relations and 
applications to new physics beyond the SM will be 
presented elsewhere\cite{10}.

The branching ratios for the decays in the relations in eqs. (\ref{pp}) 
and (\ref{pv})  
are in the range of $10^{-5} \sim 10^{-6}$ 
and some of the decay modes of $B_{u,d}$ 
have been measured. 
Although none of the $B_s$ decay modes involved have been measured, 
they are expected to be
measured at near future hadron colliders experiments 
such as CDF, D0, HERA-B, 
LHC-B and BTeV. With
$10^{8}$ mesons for each of the $B_{u,d,s}$, most of the relations 
discussed can be tested at a level better
than 10\%. Combined studies, such as taking the sum of some 
of the branching ratios of the
$B_{u,d}$ decays on the left-hand side of the equalities 
and the corresponding decay modes for $B_s$ decays on the
right-hand side, can also be carried out to increase the 
statistics and have earlier tests.

We emphasis that the relations discussed here provide 
electroweak model independent tests for
SU(3) symmetry in hadronic B decays and therefore important 
information about 
the QCD hadronic dynamics. Only when these relations 
are established, one can have confidence in using
SU(3) relations to extract important parameters, 
such as the phase angle $\gamma$, in the
SM and to test models beyond the SM using other 
relations predicted by SU(3) symmetry which are 
electroweak model dependent\cite{7}. For a full consistent test
of SU(3) symmetry,
one also needs to make sure that annihilations are indeed small.
It is therefore also important to measure some pure annihilation 
decays such as 
$B_d \to K^+ K^-, K^{*+} K^{*-}$, $K^+ K^{*-}, K^- K^{*+}$, 
$B_s \to \pi^+\pi^-, \pi^0\pi^0$, $\rho^+ \rho^-, 
\rho^0 \rho^0, \pi^+ \rho^-, \pi^-\rho^+, \pi^0 \rho^0$ 
to good precisions. 
We urge our experimental colleagues to measure these decays and
the decays in eqs. (\ref{pp}) and
(\ref{pv}) to carry out electroweak model independent tests for
SU(3) symmetry in B decays.

This work was supported in part by NSC of R.O.C. 
under grant number NSC 89-2112-M-002-016. We
thank J.-Q. Shi for discussions.

\begin{table}[ht]
\caption{
SU(3) invariant amplitudes for $B\to PV$ decays
in models beyond SM. 
The factors $a_i^{V,M}$ and $b_i^{V,M}$ are defined through
$P^{\bar q \bar dd, \bar q \bar s s} = 
\sum_i [a_i^V A_i^V + a_i^M A_i^M + b_i^V C_i^V 
+ b_i^M C^M_i]$ as in Ref.[6].
It is understood that $A_i$ and $C_i$ associated with each 
type of operators are different.}
\small{
\begin{tabular}{|c|c|c|c|c|c|c|c|c|c|c|c|}
Decay modes&$a_{\bar 3}$&$a_6^V$&$a_6^M$&
$a_{\overline{15}}^V$&$a_{\overline{15}}^M$
&$b_{\bar 3}^V$&$b_{\bar 3}^M$&$b_6^V$&$b_6^M$&
$b_{\overline{15}}^V$&$b_{\overline{15}}^M$\\
\hline
$P^{\bar d \bar dd}_{B_d\to \pi^-\rho^+}$&1&0&0&-2&4&1&0&0&0&-2&0  \\
$P^{\bar d\bar dd}_{B_s\to \pi^- K^{*+}}$&0&0&0&0&-2&1&0&0&0&-2&0   \\
$P^{\bar d\bar s s}_{B_d \to \pi^-\rho^+}$&1&-1&0&-1&-2&1&0&1&0&-1&0  \\
$P^{\bar d \bar s s}_{B_s \to \pi^- K^{*+}}$&0&0&-1&0&3&1&0&1&0&-1&0 \\
\hline
$P^{\bar d \bar dd}_{B_d\to \pi^+\rho^-}$&1&0&0&4&-2&0&1&0&0&0&-2\\
$P^{\bar d\bar dd}_{B_s\to K^+ \rho^-}$&0&0&0&-2&0&0&1&0&0&0&-2\\
$P^{\bar d\bar s s}_{B_d \to \pi^+\rho^-}$&1&0&-1&-2&-1&0&1&0&1&0&-1\\
$P^{\bar d \bar s s}_{B_s \to K^+ \rho^-}$&0&-1&0&3&0&0&1&0&1&0&-1\\
\hline
$P^{\bar s \bar ss}_{B_d\to \rho^+ K^{-}}$&0&0&0&0&-2&1&0&0&0&-2&0\\
$P^{\bar s\bar ss}_{B_s\to K^- K^{*+}}$&1&0&0&-2&4&1&0&0&0&-2&0\\
$P^{\bar s\bar dd}_{B_d \to \rho^+ K^{-}}$&0&0&-1&0&3&1&0&1&0&-1&0\\
$P^{\bar s \bar dd}_{B_s \to K^- K^{*+}}$&1&-1&0&-1&-2&1&0&1&0&-1&0\\
\hline
$P^{\bar s \bar ss}_{B_d\to \pi^+ K^{*-}}$&0&0&0&-2&0&0&1&0&0&0&-2\\
$P^{\bar s\bar ss}_{B_s\to K^+ K^{*-}}$&1&0&0&4&-2&0&1&0&0&0&-2\\
$P^{\bar s\bar dd}_{B_d \to \pi^+ K^{*-}}$&0&-1&0&3&0&0&1&0&1&0&-1\\
$P^{\bar s \bar dd}_{B_s \to K^+ K^{*-}}$&1&0&-1&-2&-1&0&1&0&1&0&-1\\
\end{tabular}
}
\end{table}

\end{document}